\documentclass[12pt,onecolumn,draftcls]{IEEEtran}

\usepackage{graphicx}
\usepackage{amssymb}
\usepackage{verbatim}
\usepackage{graphicx}
\usepackage{amsmath}
\usepackage{wrapfig}
\usepackage{algorithm}
\usepackage{algorithmic}
\usepackage{slashbox}

\begin{document}
\title{SPARLS: A Low Complexity Recursive $\mathcal{L}_1$-Regularized Least Squares Algorithm}
\author{Behtash~Babadi,
        Nicholas~Kalouptsidis
        and~Vahid~Tarokh
\thanks{B. Babadi and V. Tarokh are with the School of Engineering and Applied Sciences, Harvard University, Cambridge,
MA, 02138. (e-mails: \{behtash , vahid\}@seas.harvard.edu)}
\thanks{N. Kalouptsidis is with the Department of Informatics and Telecommunications, National and Kapodistrian University of Athens, Athens, Greece (e-mail: kalou@di.uoa.gr)}}
\maketitle
\begin{abstract}
We develop a Recursive $\mathcal{L}_1$-Regularized Least Squares (SPARLS) algorithm for the estimation of a sparse tap-weight vector in the adaptive filtering setting. The SPARLS algorithm exploits noisy observations of the tap-weight vector output stream and produces its estimate using an Expectation-Maximization type algorithm. Simulation studies in the context of channel estimation, employing multi-path wireless channels, show that the SPARLS algorithm has significant improvement over the conventional widely-used Recursive Least Squares (RLS) algorithm, in terms of both mean squared error (MSE) and computational complexity.
\end{abstract}


\section{Introduction}
Adaptive filtering is an important part of statistical signal processing, which is highly appealing in estimation problems based on streaming data in environments with unknown statistics \cite{haykin}. In particular, it is widely used for echo cancellation in speech processing systems and for equalization or channel estimation in wireless systems.\\

A wide range of signals of interest admit sparse representations. Furthermore various input output systems are described by sparse models. For example, the multi-path wireless channel has only a few significant components \cite{nowak3}. Other examples include echo components of sound in indoor environments and natural images. However, the conventional adaptive filtering algorithms, such as Least Mean Squares (LMS) and Recursive Least Squares (RLS) algorithms, which are widely used in practice, do not exploit the underlying sparseness in order to improve the estimation process.\\

There has been a lot of focus on the estimation of sparse signals based on noisy observations among the researchers in the fields of signal processing and information theory (Please see \cite{mehmet}, \cite{candes1}, \cite{candes2}, \cite{donoho1}, \cite{nowak}, \cite{tropp} and \cite{wain}). Although the above-mentioned works contain fundamental theoretical results, most of the proposed estimation algorithms are not tailored to time varying environments with real time requirements; they suffer from high complexity and are not appropriate for implementation purposes.\\

Recently, Bajwa et. al \cite{nowak3} used the Dantzig Selector (presented by Candes and Tao \cite{candes2}) and Least Squares (LS) estimates for the problem of sparse channel sensing. Although the Dantzig Selector and the LS method produce sparse estimates with improved MSE, they do not exploit the sparsity of the underlying signal in order to reduce the computational complexity. Moreover, they are not appropriate for the setting of streaming data.\\

In this paper, we introduce a Recursive $\mathcal{L}_1$-Regularized Least Squares (SPARLS) algorithm for adaptive filtering setup. The SPARLS algorithm is based on an Expectation-Maximization (EM) type algorithm presented in \cite{nowak2} and produces successive improved estimates based on streaming data. Simulation studies show that the SPARLS algorithm significantly outperforms the RLS algorithm both in terms of MSE and computational complexity for static and time-varying sparse signals. In particular, for estimating a time-varying Rayleigh fading wireless channel with 5 nonzero coefficients, the SPARLS gains about 7dB over the RLS algorithm in MSE and has about 70$\%$ less computational complexity.\\

The outline of the paper is as follows: We will introduce the notation in Section \ref{notation}. The adaptive filtering setup is discussed in Section \ref{adaptive_filtering_setup}. We will explain the regularized cost function in Section \ref{regularized_cost_function}. An efficient algorithm to optimize the regularized cost function, namely Low-Complexity Expectation Maximization (LCEM) is introduced in Section \ref{LCEM_alg}. We will formally define the SPARLS algorithm along with complexity analysis and related discussions in Section \ref{SPARLS_algorithm}. Simulation studies are presented in Section \ref{simulation_studies}, followed by conclusion in Section \ref{conclusion}.

\section{Notation}\label{notation}

Let $\textbf{x}$ be a vector in $\mathbb{C}^{M}$. We define the $\mathcal{L}_0$ quasi-norm of $\textbf{x}$ as follows:
\begin{equation}
\| \textbf{x} \|_0 = | \{ x_i | x_i \neq 0 \} |
\end{equation}
A vector $\textbf{x} \in \mathbb{C}^{M}$ is called sparse, if $\| \textbf{x} \|_0 \ll M$. Let $\textbf{A}$ be a matrix in $\mathbb{C}^{N \times M}$ and $\mathcal{J} \subseteq \{1,2,\cdots,M\}$ be an index set. We denote the sub-matrix of $\textbf{A}$ with columns corresponding to the index set $\mathcal{J}$ by $\textbf{A}_{\mathcal{J}}$. Similarly, we denote the sub-vector of $\textbf{x} \in \mathbb{C}^{M}$ corresponding to the index set $\mathcal{J}$ by $\textbf{x}_{\mathcal{J}}$. We denote the conjugate transpose of $\textbf{A} \in \mathbb{C}^{N \times M}$ and $\textbf{x} \in \mathbb{C}^M$ by $\textbf{A}^*$ and $\textbf{x}^*$, respectively. We also define the element-wise magnitude and signum operators as follows:
\begin{equation}
|\textbf{x}| := [|x_1|, |x_2|, \cdots, |x_M|]^T
\end{equation}
and
\begin{equation}
\operatorname{sgn}(\textbf{x}) := [\operatorname{sgn}(x_1), \operatorname{sgn}(x_2), \cdots, \operatorname{sgn}(x_M)]^T
\end{equation}
for $\textbf{x} \in \mathbb{R}^M$, where
\begin{equation}
\operatorname{sgn}(x_i) :=
\left\{ {\begin{array}{*{20}c}
   {1} &  {x_i \ge 0}  \\
   {-1} & {x_i <0}  \\
\end{array}} \right.
\end{equation}

For $\textbf{x}, \textbf{y} \in \mathbb{C}^M$, we define the element-wise multiplication as follows:
\begin{equation}
\textbf{x}\cdot\textbf{y} := [x_1y_1, x_2y_2, \cdots, x_My_M]^T
\end{equation}
For any $\textbf{x} \in \mathbb{R}^M$, we define
\begin{equation}
\textbf{x}_+ := [(x_1)_+, (x_2)_+, \cdots,(x_M)_+]^T
\end{equation}
where
\begin{equation}
(x_i)_+ := \max(x_i,0)
\end{equation}

Finally, we define the all-one vector in $\mathbb{R}^M$ as
\begin{equation}
\textbf{1} := [1, 1, \cdots,1]^T.
\end{equation}

\section{Adaptive Filtering Setup}\label{adaptive_filtering_setup}

\subsection{Canonical Adaptive Filtering Setup}

\begin{figure} \label{model}
\begin{center}
    \includegraphics[bb=73pt 250pt 551pt 540pt, scale=0.8]{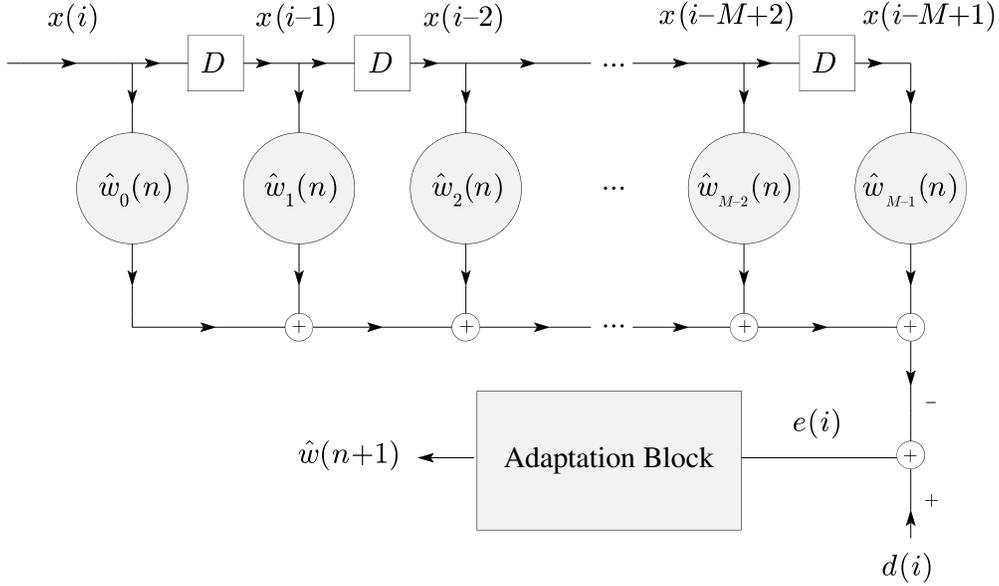}
\end{center}
\caption{Adaptive filtering model}
\end{figure}

Consider the conventional adaptive filtering setup, consisting of a transversal filter followed by an adaptation block (Fig. 1). The tap-input vector at time $i$ is defined by
\begin{equation}
\textbf{x}(i) := [x(i), x(i-1),\cdots, x(i-M+1)]^T
\end{equation}
where $x(k)$ is the input at time $k$, $k=1,\cdots,n$. The tap-weight vector at time $n$ is defined by
\begin{equation}
\hat{\textbf{w}}(n) := [\hat{w}_0(n), \hat{w}_1(n),\cdots, \hat{w}_{M-1}(n)]^T
\end{equation}
Note that the tap-weight vector is assumed to be constant during the observation time $1 \le i \le n$. The output of the filter at time $i$ is defined by
\begin{equation}
y(i) := \hat{\textbf{w}}^*(n) \textbf{x}(i)
\end{equation}
The tap-weight vector $\hat{\textbf{w}}(n)$ is updated by the adaptation block in order to optimize a certain cost function. Let $d(i)$ be the desired output of the filter at time $i$. We can define the instantaneous error of the filter by
\begin{equation}
e(i) := d(i) - y(i) = d(i) - \hat{\textbf{w}}^*(n) \textbf{x}(i)
\end{equation}
The operation of the adaptation block at time $n$ can therefore be stated as the following optimization problem:
\begin{equation}\label{canonical}
\min_{\hat{\textbf{w}}(n)} \mbox{ } \mbox{ } \mbox{ } f\big(e(1),e(2),\cdots,e(n)\big)
\end{equation}
where $f \ge 0$ is a certain cost function. The adaptation block exploits $e(i)$, $i=1,2,\cdots,n$ in order to adjust $\hat{\textbf{w}}(n)$. Note that there is no constraint on the desired output $d(i)$ and $f$ so far. With appropriate choices for $d(i)$ and $f$, one can recast problems such as channel estimation, echo cancellation and equalization in the canonical form given by Eq. (\ref{canonical}).\\

In particular, if $d(i)$ is generated by an unknown tap-weight $\textbf{w}(n)$, \emph{i.e.}, $d(i) = {\textbf{w}}^*(n) \textbf{x}(i)$, with an appropriate choice of $f$, one can possibly obtain a good approximation to $\textbf{w}(n)$ by solving the optimization problem given in (\ref{canonical}). This is, in general, an estimation problem and is the topic of interest in this paper\footnote[1]{Our discussion will focus on single channel complex valued signals. The extension to the multi-variable case presents no difficulties.}.

\subsection{Examples of Conventional Cost Functions}
There are various choices for $f$ which give rise to certain approximations of the unknown vector $\textbf{w}(n)$. For example, one can choose $f$ to be
\begin{equation}
f_{LMS}\big(e(1),e(2),\cdots,e(n)\big) := |e(n)|^2
\end{equation}
which is the squared error at time $n$. Solving the optimization problem given in Eq. (\ref{canonical}) for $f_{LMS}$ will result in the well-known Least Mean Squares (LMS) algorithm \cite{haykin}. Note that $f_{LMS}$ is memoryless, \emph{i.e.}, it only depends on $e(n)$. Another choice for $f$ can be
\begin{equation}
f_{M}\big(e(1),e(2),\cdots,e(n)\big) := \sum_{i=1}^n \beta(i,n) |e(i)|^2
\end{equation}
where $\beta(i,n)$, $i=1,2,\cdots,n$ are non-negative constants, introducing memory to the cost function. Two useful weighting sequences are the sliding window, where $\beta(i,n)$ is zero prior to a positive integer $l<n$, and the exponentially decaying sequence
\begin{equation}
\beta(i,n) = \lambda^{n-i}
\end{equation}
for $i=1,2,\cdots,n$ and $\lambda$ a non-negative constant. Using the latter weighting sequence, the cost function takes the form
\begin{equation}\label{f_RLS}
f_{RLS}\big(e(1),e(2),\cdots,e(n)\big) := \sum_{i=1}^n \lambda^{n-i} |e(i)|^2 .
\end{equation}
The parameter $\lambda$ is commonly referred to as \emph{forgetting factor}. The solution to the optimization problem in Eq. (\ref{canonical}) with $f_{RLS}$ gives rise to the well-known Recursive Least Squares (RLS) algorithm. It is well known \cite{haykin} that RLS enjoys faster convergence than LMS (where $\lambda=0$). The cost function $f_{RLS}$ given in (\ref{f_RLS}) corresponds to a least squares identification problem. Let
\begin{equation}
\textbf{D}(n) :=
\left( {\begin{array}{*{20}c}
   {\lambda ^{n - 1} } &  \cdots  & 0 & 0  \\
    \vdots  & {\lambda ^{n - 2} } &  \vdots  & 0  \\
   0 &  \cdots  &  \ddots  &  \vdots   \\
   0 & 0 &  \cdots  & 1  \\
\end{array}} \right) ,
\end{equation}
\begin{equation}
\textbf{d}(n) := [d^*(1), d^*(2), \cdots, d^*(n)]^T
\end{equation}
and $\textbf{X}(n)$ be an $n \times M$ matrix whose $i$th row is $\textbf{x}^*(i)$, \emph{i.e.},
\begin{equation}
\textbf{X}(n) :=
\left( {\begin{array}{*{20}c}
   {x^*(1)} &  \cdots  & 0 & 0  \\
    \vdots  &  \vdots  &  \vdots  &  \vdots   \\
   {x^*(n - 1)} & {x^*(n - 2)} &  \cdots  & {x^*(n - M)}  \\
   {x^*(n)} & {x^*(n - 1)} &  \cdots  & {x^*(n - M + 1)}  \\
\end{array}} \right).
\end{equation}
The cost function can be written in the following form:
\begin{equation}
f_{RLS}\big(e(1),e(2),\cdots,e(n)\big) = \| {\textbf{D}^{1/2}(n)} \textbf{d}(n) - {\textbf{D}^{1/2}(n)} \textbf{X}(n) \hat{\textbf{w}}(n) \|_2^2
\end{equation}
where $\textbf{D}^{1/2}(n)$ is a diagonal matrix with entries $D^{1/2}_{ii}(n) := \sqrt{{D}_{ii}(n)}$. Thus, the solution to the optimization problem with the cost function $f_{RLS}$ can be expressed in terms of the following normal equations \cite{haykin}:
\begin{equation}
\boldsymbol\Phi(n) \hat{\textbf{w}}(n) = \textbf{z}(n)
\end{equation}
where
\begin{equation}
\boldsymbol\Phi(n) := \textbf{X}^*(n) \textbf{D}(n) \textbf{X}(n) = \sum_{i=1}^n \lambda^{n-i} \textbf{x}(i) \textbf{x}^*(i)
\end{equation}
and
\begin{equation}
\textbf{z}(n) := \textbf{X}^*(n) \textbf{D}(n) \textbf{d}(n)  = \sum_{i=1}^n \lambda^{n-i} \textbf{x}(i)d^*(i)
\end{equation}

\section{Regularized Cost Function}\label{regularized_cost_function}

\subsection{Noisy Observations}

The canonical form of the problem typically assumes that the input-output sequences are generated by a time varying system with parameters represented by $\textbf{w}(n)$. In most applications however, stochastic uncertainties are also present. Thus a more pragmatic data generation process is described by the noisy model
\begin{equation}
d(i) = {\textbf{w}}^*(n) \textbf{x}(i) + \eta(i)
\end{equation}
where $\eta(i)$ is the observation noise. Note that $\textbf{w}(n)$ reflects the true parameters which vary with time in a piecewise constant manner. The noise will be assumed to be i.i.d. Gaussian, i.e., $\eta(i) \thicksim \mathcal{N}(0,\sigma^2)$. The estimator has only access to the streaming data $x(i)$ and $d(i)$.

\subsection{Estimation of Sparse Vectors}

A wide range of interesting estimation problems deal with the estimation of sparse vectors. Many signals of interest can naturally be modeled as sparse. For example, the wireless channel usually has a few significant multi-path components. One needs to estimate such signals for various purposes. Suppose that $\|\textbf{w}(n)\|_0 = L \ll M$.
A sparse approximation to $\textbf{w}(n)$ can be obtained by solving the following optimization problem:
\begin{equation}\label{l_0}
\min_{\hat{\textbf{w}}(n)} \| \hat{\textbf{w}}(n) \|_0 \mbox{ } \mbox{ } \mbox{s.t.} \mbox{ } \mbox{ } f\big(e(1),e(2),\cdots,e(n)\big) \le \epsilon
\end{equation}
where $\epsilon$ is a positive constant controlling the cost error in (\ref{canonical}). The above optimization problem is computationally intractable. A considerable amount of recent research in statistical signal processing is focused on efficient estimation methods for estimating an unknown sparse vector based on noiseless/noisy observations (Please see \cite{candes1}, \cite{candes2}, \cite{donoho1}, \cite{nowak4} and \cite{nowak}). In particular, convex relaxation techniques provide a viable alternative, whereby the $\mathcal{L}_0$ quasi-norm in (\ref{l_0}) is replaced by the convex $\mathcal{L}_1$ norm so that (\ref{l_0}) becomes
\begin{equation}\label{l_1}
\min_{\hat{\textbf{w}}(n)} \| \hat{\textbf{w}}(n) \|_1 \mbox{ } \mbox{ } \mbox{s.t.} \mbox{ } \mbox{ } f\big(e(1),e(2),\cdots,e(n)\big) \le \epsilon
\end{equation}
A convex problem results when $f$ is convex, as in the RLS case. The Lagrangian formulation shows that if $f=f_{RLS}$, the optimum solution can be equivalently derived from the following optimization problem
\begin{equation}\label{convex}
\min_{\hat{\textbf{w}}(n)} \mbox{  } \bigg\{ \frac{1}{2 \sigma^2} \big\| \textbf{D}^{1/2}(n) \textbf{d}(n) - \textbf{D}^{1/2}(n) \textbf{X}(n) \hat{\textbf{w}}(n) \big \|_2^2 + \gamma \| \hat{\textbf{w}}(n) \|_1 \bigg\}
\end{equation}
$\gamma$ represents a trade off between estimation error and sparsity of the parameter coefficients. Sufficient as well as necessary conditions for the existence and uniqueness of a global minimizer are derived in \cite{tropp}. These conditions require that the input signal must be properly chosen so that the matrix $\textbf{D}^{1/2}(n) \textbf{X}(n)$ is sufficiently incoherent. Suitable probing signals for exact recovery in a multi-path environment are analyzed in \cite{nowak3}.


\section{Low-Complexity Expectation Maximization Algorithm}\label{LCEM_alg}

The convex program in Eq. (\ref{convex}) can be solved with the conventional convex programming methods. Here, we adopt an efficient solution presented by Nowak \cite{nowak2} in the context of Wavelet-based image restoration, which we will modify to an online and adaptive setting. Consider the noisy observation model:
\begin{equation}
\textbf{d}(n) = \textbf{X}(n) \textbf{w}(n) + {\boldsymbol\eta}(n) .
\end{equation}
where $\boldsymbol\eta(n) \thicksim \mathcal{N}(0,\sigma^2 \textbf{I})$, with the following cost function
\begin{eqnarray}\label{convex2}
&& \frac{1}{2 \sigma^2} \big\| \textbf{D}^{1/2}(n) \textbf{d}(n) - \textbf{D}^{1/2}(n) \textbf{X}(n) \hat{\textbf{w}}(n) \big\|_2^2 + \gamma \| \hat{\textbf{w}}(n) \|_1\\
\nonumber &=& \frac{1}{2 \sigma^2} \Big(\textbf{d}(n) - \textbf{X}(n) \hat{\textbf{w}}(n)\Big)^* \textbf{D}(n) \Big(\textbf{d}(n) - \textbf{X}(n) \hat{\textbf{w}}(n)\Big)+ \gamma \| \hat{\textbf{w}}(n) \|_1
\end{eqnarray}
If we consider the alternative observation model:
\begin{equation}\label{model}
\textbf{d}(n) = \textbf{X}(n) \textbf{w}(n) + {\boldsymbol\xi}(n) .
\end{equation}
with $\boldsymbol\xi(n) \thicksim \mathcal{N}(0,\sigma^2 \textbf{D}^{-1}(n))$, the convex program in Eq. (\ref{convex}) can be identified as the following Maximum Likelihood (ML) problem:
\begin{equation}\label{like}
\max_{\textbf{w}(n)} \mbox{  } \Big\{\log p(\textbf{d}(n)|\textbf{w}(n)) - \gamma \| \textbf{w}(n) \|_1\Big\}
\end{equation}
where $p(\textbf{d}(n)|\textbf{w}(n)) := \mathcal{N}(\textbf{X}(n)\textbf{w}(n), \sigma^2 \textbf{D}^{-1}(n))$. This ML problem is in general hard to solve. The clever idea of \cite{nowak2} is to decompose the noise vector $\boldsymbol\xi(n)$ in order to divide the optimization problem into a denoising and a filtering problem. We adopt the same method with appropriate modifications for the cost function given in Eq. (\ref{like}). Consider the following decomposition for $\boldsymbol\xi(n)$:
\begin{equation}
\boldsymbol\xi(n) = \alpha \textbf{X}(n) \boldsymbol\xi_1(n) + \boldsymbol\xi_2(n)
\end{equation}
where $\boldsymbol\xi_1(n) \thicksim \mathcal{N}(0,\textbf{I})$ and $\boldsymbol\xi_2(n) \thicksim \mathcal{N}(0, \sigma^2 {\textbf{D}^{-1}(n)} - \alpha^2 \textbf{X}(n) \textbf{X}^*(n))$. We need to choose $\alpha^2 \le \sigma^2/s_1$, where $s_1$ is the largest eigenvalue of $\textbf{X}(n) \textbf{X}^*(n)$, in order for $\boldsymbol\xi_2(n)$ to have a positive semi-definite covariance matrix. We can therefore rewrite the model in Eq. (\ref{model}) as
\begin{equation}
\left\{ {\begin{array}{*{20}l}
   {\textbf{v}(n) = \textbf{w}(n) + \alpha \boldsymbol\xi_1(n)}  \\
   {\textbf{d}(n) =  \textbf{X}(n) \textbf{v}(n) +  \boldsymbol\xi_2(n)}  \\
\end{array}} \right.
\end{equation}
The Expectation Maximization (EM) algorithm can be used to solve the ML problem of (\ref{like}), with the help of the following ML problem
\begin{equation}\label{like_easy}
\max_{\textbf{w}(n)} \mbox{  } \Big\{\log p(\textbf{d}(n), \textbf{v}(n)|\textbf{w}(n)) - \gamma \| \textbf{w}(n) \|_1\Big\} ,
\end{equation}
which is easier to solve. The $\ell$th iteration of the EM algorithm is as follows:
\begin{equation}\label{iter}
\left\{ {\begin{array}{*{20}l}
   {\textbf{r}^{(\ell)}(n) = \big(\textbf{I} - \frac{\alpha^2}{\sigma^2} \textbf{X}^*(n) \textbf{D}(n) \textbf{X}(n)\big) \hat{\textbf{w}}^{(\ell)}(n) + \frac{\alpha^2}{\sigma^2} \textbf{X}^*(n)\textbf{D}(n)\textbf{d}(n)}  \\
   {\hat{\textbf{w}}^{(\ell+1)}(n) = \operatorname{sgn}\Big(\textbf{r}^{(\ell)}(n)\Big)\cdot\Big(|\textbf{r}^{(\ell)}(n)| - \gamma \alpha^2 \textbf{1}\Big)_+}  \\
\end{array}} \right.
\end{equation}
The function $\operatorname{sgn}(x)\big(|x| - \gamma \alpha^2\big)_+$ is denoted by \emph{soft thresholding function} and is plotted in Fig. 2.

\begin{figure} \label{st}
\begin{center}
    \includegraphics[bb=65pt 211pt 547pt 574pt, scale=.5]{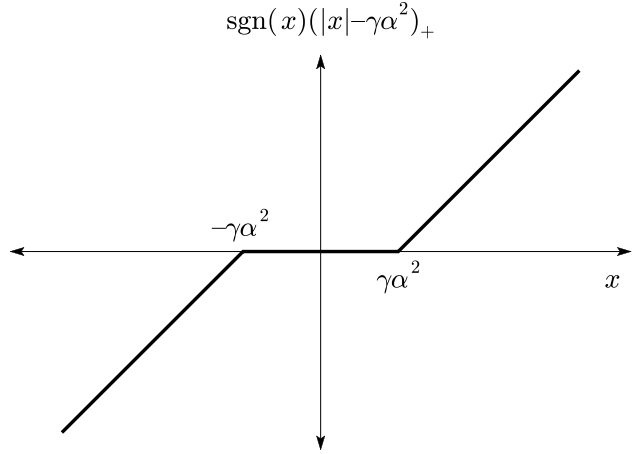}
\end{center}
\caption{Soft thresholding function}
\end{figure}

It is known that the EM algorithm given by Eq. (\ref{iter}) converges \cite{vantrees}. Note that the soft thresholding function tends to decrease the support of the estimate $\hat{\textbf{w}}(n)$, since it will shrink the support to those elements whose absolute value is greater than $\gamma \alpha^2$. We can use this observation to express the double iteration given in Eq. (\ref{iter}) in a low complexity fashion. Let $\mathcal{I}^{(\ell)}$ be the support of $\textbf{r}^{(\ell)}(n)$ at the $\ell$th iteration. Let
\begin{equation}
\left\{ {\begin{array}{*{20}l}
   {\mathcal{I}_+^{(\ell)} := \{ i : {r}_i^{(\ell)}(n) > \gamma \alpha^2 \} \subseteq \mathcal{I}^{(\ell)}}  \\
   {\mathcal{I}_-^{(\ell)} := \{ i : {r}_i^{(\ell)}(n) < -\gamma \alpha^2 \} \subseteq \mathcal{I}^{(\ell)}} \\
\end{array}} \right. ,
\end{equation}
\begin{equation}
\textbf{B}(n) := \textbf{I} - \frac{\alpha^2}{\sigma^2} \textbf{X}^*(n) \textbf{D}(n) \textbf{X}(n) ,
\end{equation}
\begin{equation}
\textbf{s}^{(\ell)}(n) := \textbf{B}(n)  \hat{\textbf{w}}^{(\ell)}(n) ,
\end{equation}
and
\begin{equation}
\textbf{u}(n) := \frac{\alpha^2}{\sigma^2} \textbf{X}^*(n){\textbf{D}(n)}\textbf{d}(n) .
\end{equation}
Note that the second iteration in Eq. (\ref{iter}) can be written as
\begin{equation}
\hat{w}_i^{(\ell+1)}(n) =\left\{ {\begin{array}{*{20}l}
   {r_i^{(\ell)}(n)-\gamma \alpha^2} &  {i \in \mathcal{I}_+^{(\ell)}}  \\
   {r_i^{(\ell)}(n)+ \gamma \alpha^2} & {i \in \mathcal{I}_-^{(\ell)}}  \\
   {0} & {i \notin \mathcal{I}_+^{(\ell)}\cup \mathcal{I}_-^{(\ell)}}\\
\end{array}} \right.
\end{equation}
for $i=1,2,\cdots,M$. We then have
\begin{equation}
\textbf{B}(n) \hat{\textbf{w}}^{(\ell+1)}(n) = \textbf{B}_{\mathcal{I}_+^{(\ell)}}(n) \big(\textbf{r}^{(\ell)}_{\mathcal{I}_+^{(\ell)}}(n) - \gamma \alpha^2 \mathbf{1}_{\mathcal{I}_+^{(\ell)}} \big) + \textbf{B}_{\mathcal{I}_-^{(\ell)}}(n) \big(\textbf{r}^{(\ell)}_{\mathcal{I}_-^{(\ell)}}(n)  + \gamma \alpha^2 \mathbf{1}_{\mathcal{I}_-^{(\ell)}} \big)
\end{equation}
which allows us to express the EM iteration as follows:
\begin{equation}\label{LCEM}
\left\{ {\begin{array}{*{20}l}
   {\textbf{s}^{(\ell+1)}(n) = \textbf{B}_{\mathcal{I}_+^{(\ell)}}(n) \big(\textbf{s}^{(\ell)}_{\mathcal{I}_+^{(\ell)}}(n) + {\textbf{u}}_{\mathcal{I}_+^{(\ell)}}(n) - \gamma \alpha^2 \mathbf{1}_{\mathcal{I}_+^{(\ell)}} \big) + \textbf{B}_{\mathcal{I}_-^{(\ell)}}(n) \big(\textbf{s}^{(\ell)}_{\mathcal{I}_-^{(\ell)}}(n) + {\textbf{u}}_{\mathcal{I}_-^{(\ell)}}(n) + \gamma \alpha^2 \mathbf{1}_{\mathcal{I}_-^{(\ell)}} \big)
}  \\
   {\textbf{r}^{(\ell+1)}(n)} = \textbf{s}^{(\ell+1)}(n) + {\textbf{u}}(n)\\
   {\mathcal{I}_+^{(\ell+1)} = \{ i : {r}_i^{(\ell+1)}(n) > \gamma \alpha^2 \}}  \\
   {\mathcal{I}_-^{(\ell+1)} = \{ i : {r}_i^{(\ell+1)}(n) < -\gamma \alpha^2 \}}  \\
\end{array}} \right.
\end{equation}
This new set of iteration has a lower computational complexity, since it restricts the matrix multiplications to the instantaneous support of the estimate $\textbf{r}^{(\ell)}(n)$, which is expected to be close to the support of $\textbf{w}(n)$ \cite{tropp}. We denote the iterations given in Eq. (\ref{LCEM}) by Low-Complexity Expectation Maximization (LCEM) algorithm.

\section{The SPARLS Algorithm}\label{SPARLS_algorithm}

\subsection{SPARLS}
Upon the arrival of the $n$th input, $x(n)$, the LCEM algorithm computes the estimate $\hat{\textbf{w}}(n)$ given $\textbf{B}(n)$, $\textbf{u}(n)$ and $\textbf{s}^{(0)}(n)$. The LCEM algorithm is summarized in Algorithm 1. Note that the input argument $k$ denotes the number of iterations.
\begin{algorithm}{}
  \caption{$\operatorname{LCEM}\big(\textbf{B}(n), \textbf{u}(n), \textbf{s}^{(0)}, k\big)$}
  Inputs: $\textbf{B}(n)$, $\textbf{u}(n)$, $\textbf{s}^{(0)}$, $k$.\\
  Outputs: $\hat{\textbf{w}}(n)$.
  \begin{algorithmic}[1]
    \STATE $\textbf{r}^{(0)} = \textbf{s}^{(0)} + \textbf{u}(n)$.
    \STATE $\mathcal{I}_+^{(0)} = \{ i : {r}_i^{(0)}(n) > 0 \}$.
    \STATE $\mathcal{I}_-^{(0)} = \{ i : {r}_i^{(0)}(n) < 0 \}$.
    \FOR{$\ell=0,1,\cdots,k-1$}
    \STATE $\textbf{s}^{(\ell+1)}(n) = \textbf{B}_{\mathcal{I}_+^{(\ell)}}(n) \big(\textbf{s}^{(\ell)}_{\mathcal{I}_+^{(\ell)}}(n) + {\textbf{u}}_{\mathcal{I}_+^{(\ell)}}(n) - \gamma \alpha^2 \mathbf{1}_{\mathcal{I}_+^{(\ell)}} \big) + \textbf{B}_{\mathcal{I}_-^{(\ell)}}(n) \big(\textbf{s}^{(\ell)}_{\mathcal{I}_-^{(\ell)}}(n) + {\textbf{u}}_{\mathcal{I}_-^{(\ell)}}(n) + \gamma \alpha^2 \mathbf{1}_{\mathcal{I}_-^{(\ell)}} \big)$
    \STATE $\textbf{r}^{(\ell+1)}(n) = \textbf{s}^{(\ell+1)}(n) + \textbf{u}(n)$.
    \STATE $\mathcal{I}_+^{(\ell+1)} = \{ i : {r}_i^{(\ell+1)}(n) > \gamma \alpha^2 \}$.
    \STATE $\mathcal{I}_-^{(\ell+1)} = \{ i : {r}_i^{(\ell+1)}(n) < -\gamma \alpha^2 \}$.
    \ENDFOR
    \FOR{$i=1,2,\cdots,M$}
    \STATE $\hat{w}_i(n) =\left\{ {\begin{array}{*{20}l}
   {r_i^{(k)}-\gamma \alpha^2} &  {i \in \mathcal{I}_+^{(k)}}  \\
   {r_i^{(k)}+ \gamma \alpha^2} & {i \in \mathcal{I}_+^{(k)}}  \\
   {0} & {i \notin \mathcal{I}_+^{(k)} \cup \mathcal{I}_-^{(k)}}\\
\end{array}} \right.$.
    \ENDFOR
  \end{algorithmic}
  \label{LCEMA}
\end{algorithm}

Upon the arrival of the $n$th input, $\textbf{B}(n)$ and $\textbf{u}(n)$ can be obtained via the following rank-one update rules:
\begin{equation}
\left\{ {\begin{array}{*{20}l}
   {\textbf{B}(n)= \lambda \textbf{B}(n-1) - \frac{\alpha^2}{\sigma^2} \textbf{x}(n) \textbf{x}^*(n)}  \\
   {\textbf{u}(n) = \lambda \textbf{u}(n-1) + \frac{\alpha^2}{\sigma^2} d^*(n) \textbf{x}(n)}\\
\end{array}} \right.
\end{equation}

The SPARLS algorithm is formally defined in Algorithm 2. Without loss of generality, we can set the time index $n=1$ such that $x(1) \neq 0$, in order for the initialization to be well-defined.

\begin{algorithm}{}
  \caption{SPARLS}
    Inputs: $\textbf{B}(1) = \textbf{I} - \frac{\alpha^2}{\sigma^2} \textbf{x}(1) \textbf{x}^*(1)$, $\textbf{u}(1)= \frac{\alpha^2}{\sigma^2} \textbf{x}(1) d^*(1)$ and $k$.\\
    Output: $\hat{\textbf{w}}(n)$.
    \begin{algorithmic}[1]
    \FORALL{Input $x(n)$}
    \STATE $\textbf{B}(n)= \lambda \textbf{B}(n-1) - \frac{\alpha^2}{\sigma^2} \textbf{x}(n) \textbf{x}^*(n)$.
    \STATE $\textbf{u}(n) = \lambda \textbf{u}(n-1) + \frac{\alpha^2}{\sigma^2} d^*(n) \textbf{x}(n)$.
    \STATE Run $\operatorname{LCEM}\big(\textbf{B}(n), \textbf{u}(n), \textbf{B}(n)\hat{\textbf{w}}(n), k\big)$.
    \STATE Update $\hat{\textbf{w}}(n)$.
    \ENDFOR
  \end{algorithmic}
  \label{SPARLS}
\end{algorithm}

\subsection{Complexity Analysis}

The LCEM algorithm requires $M\big(|\mathcal{I}_+^{(\ell)}| + |\mathcal{I}_-^{(\ell)}|\big)$ multiplications at the $\ell$th iteration. Thus, for a total of $k$ iterations, the number of multiplications will be $k M N$, where
\begin{equation}
N := \frac{1}{k} \sum_{\ell=0}^{k-1} \big(|\mathcal{I}_+^{(\ell)}| + |\mathcal{I}_-^{(\ell)}|\big)
\end{equation}
For a sparse signal $\textbf{w}(n)$, one expects to have $N \approx \mathcal{O}(\| \textbf{w}(n) \|_0) = \mathcal{O}(L)$. Therefore, the complexity of the LCEM algorithm is roughly of the order $\mathcal{O}(k L M)$. Simulation results show that a single LCEM iteration ($k=1$) is sufficient for the SPARLS algorithm to result in significant gains in terms of both MSE and computational complexity. Note that the Recursive Least Squares (RLS) algorithm requires $\mathcal{O}(M^2)$ multiplications, which clearly has higher complexity compared to the SPARLS.

\subsection{Discussion of the SPARLS Algorithm}\label{disc}

The parameter $\alpha$ in the SPARLS algorithm must be chosen such that $\alpha^2 \le \sigma^2/s_1$, where $s_1$ is the largest eigenvalue of $\textbf{X}^*(n) \textbf{X}(n)$. For large $n$, the eigenvalues of $\textbf{X}^*(n) \textbf{X}(n)$ will all tend to $1$, given $x(i) \thicksim \mathcal{N}(0,1/M)$ for $i=1,2,\cdots,n$. Therefore, by choosing $\alpha = \sigma /2$, the condition of $\alpha^2 \le \sigma^2/s_1$ is satisfied with overwhelming probability.\\

The parameter $\gamma$ is an additional degree of freedom which controls the trade-off between sparseness of the output (computational complexity) and the MSE. For very small values of $\gamma$, the SPARLS algorithm coincides with the RLS algorithm. For very large values of $\gamma$, the output will be the zero vector. Thus, there are intermediate values for $\gamma$ which result in low MSE and sparsity level which is desired.\\

The parameter $\gamma$ can be fine-tuned according to the application we are interested in. For example, for estimating the wireless multi-path channel, $\gamma$ can be optimized with respect to the number of channel taps (sparsity), temporal statistics of the channel and noise level via exhaustive simulations or experiments. Note that $\gamma$ can be fine-tuned offline for a certain application. There are also some heuristic methods for choosing $\gamma$ which are discussed in \cite{nowak2}.\\

\section{Simulation Studies}\label{simulation_studies}

We consider the estimation of a sparse multi-path wireless channel generated by the Jake's model \cite{jakes}. In the Jake's model, each component of the tap-weight vector is a sample path of a Rayleigh random process with autocorrelation function given by
\begin{equation}
R(n)=\operatorname{J_0}(2 \pi n f_d T_s)
\end{equation}
where $\operatorname{J_0}(\cdot)$ is the zeroth order Bessel function, $f_d$ is the Doppler frequency shift and $T_s$ is the
channel sampling interval. The dimensionless parameter $f_d T_s$ gives a measure of how fast each tap is changing over time. Note that the case $f_d = 0$ corresponds to a constant tap-weight vector. Thus, the Jake's model covers constant tap-weight vectors as well. For the purpose of simulations, $T_s$ is normalized to 1.\\

In order to compare the performance of the SPARLS and RLS algorithms, we first need to optimize the RLS algorithm for the given time-varying channel. By exhaustive simulations, the optimum forgetting factor, $\lambda$, of the RLS algorithm can be obtained for various choices of SNR and $f_d$. The optimal values of $\lambda$ for several choices of SNR and $f_d$ are summarized in Table 1.

\begin{table}
\caption{Optimal values of $\lambda$, given $\sigma^2$ and $f_d$, for the RLS algorithm. Each entry has an error of $\pm 0.01$.}
\begin{center}
  \begin{tabular}{|l||*{6}{c|}}
    \hline
    {\backslashbox{$\sigma^2$}{$f_d$}} & 0 & 0.0001 & 0.0005 & 0.001 & 0.005 & 0.01\\ \hline & \\[-2em]\hline
    0.0001 & 0.98 & 0.95 & 0.95 & 0.99& 0.99 & 0.99\\ \hline
    0.0005 & 0.99 & 0.97 & 0.98& 0.99 & 0.99 & 0.99 \\ \hline
    0.001 &  0.99 & 0.97  & 0.98  & 0.99 & 0.99 & 0.99\\ \hline
    0.005 & 0.99 & 0.99 & 0.99 & 0.99& 0.99 & 0.99 \\ \hline
    0.01 & 0.99 & 0.99 & 0.99 & 0.99& 0.99& 0.99\\ \hline
    0.05 & 0.99 & 0.99 & 0.99& 0.99& 0.99& 0.99\\
    \hline
  \end{tabular}
\end{center}
\end{table}

As for the SPARLS algorithm, we perform a partial optimization as follows: we use the values of Table 1 for $\lambda$ and optimize over $\gamma$ with exhaustive simulations. The optimal values of $\gamma$ are summarized in Table 2. Note that with such choices of parameters $\lambda$ and $\gamma$, we are comparing a near-optimal parametrization of SPARLS with the optimal parametrization of RLS. The performance of the SPARLS can be further enhanced by simultaneous optimization over both $\lambda$ and $\gamma$.

\begin{table}
\caption{Optimal values of $\gamma$, given $\sigma^2$, $\lambda$ and $f_d$, for the SPARLS algorithm. Each entry has an error of $\pm 5$.}
\begin{center}
  \begin{tabular}{ |l || c | c | c | c | c | c |}
    \hline
    {\backslashbox{$\sigma^2$}{$f_d$}} & 0 & 0.0001 & 0.0005 & 0.001 & 0.005 & 0.01\\ \hline & \\[-2em]\hline
    0.0001 & 100 & 100 & 100 & 100  & 100 & 100\\ \hline
    0.0005 & 45  & 40 & 40 & 60 & 50 & 50\\ \hline
    0.001 & 30 &  25& 30& 25 & 25 &25 \\ \hline
    0.005 & 15 & 15 &10 &10 & 10 & 10\\ \hline
    0.01 & 10 & 10 & 5 & 5 & 5 &5  \\ \hline
    0.05 & 5 & 5 & 3 &2 &2 &2\\
    \hline
  \end{tabular}
\end{center}
\end{table}

We compare the performance of the SPARLS and RLS with respect to two performance measures. The first measure is the MSE defined as
\begin{equation}
\mbox{MSE} := \frac{\mathbb{E}\{ \|\hat{\textbf{w}} - \textbf{w}\|_2^2 \}}{\mathbb{E} \{ \| \textbf{w} \|_2^2 \}}
\end{equation}
where the averaging is carried out by 50000 Monte Carlo samplings. The number of samples was chosen large enough to ensure that the uncertainty in the measurements is less than $1\%$. The second measure is the computational complexity ratio (CCR) which is defined by
\begin{equation}
\mbox{CCR} := \frac{\mbox{average number of multiplications for SPARLS}}{\mbox{average number of multiplications for RLS}}
\end{equation}
In all simulations the input data $x(i)$ is i.i.d. and distributed according to $\mathcal{N}(0,1/M)$. The SNR is also defined as $\mathbb{E} \{ \| \textbf{w} \|_2^2\} / \sigma^2$, where $\sigma^2$ is the variance of the Gaussian zero-mean observation noise. The locations of the nonzero elements of the tap-weight vector are randomly chosen in the set $\{1,2,\cdots,M\}$ and the SPARLS algorithm has no knowledge of these locations. Also, all the simulations are done with $k=1$, \emph{i.e.}, a single LCEM iteration per new data. Finally, a choice of $\alpha=\sigma/2$ has been used (Please see Section \ref{disc}).\\

%

Figures 3 and 4 show the mean squared error and computational complexity ratio of the RLS and SPARLS algorithms for $f_d = 0, 0.0001, 0.0005, 0.001, 0.005$ and $0.01$, with $L=5$ and $M=100$, respectively. The SPARLS algorithm outperforms the RLS algorithm with about 7 dB gain in the MSE performance. Moreover, the computational complexity of the SPARLS is about 70$\%$ less than that of RLS.

\begin{figure}
\begin{center}
    \includegraphics[width=6in]{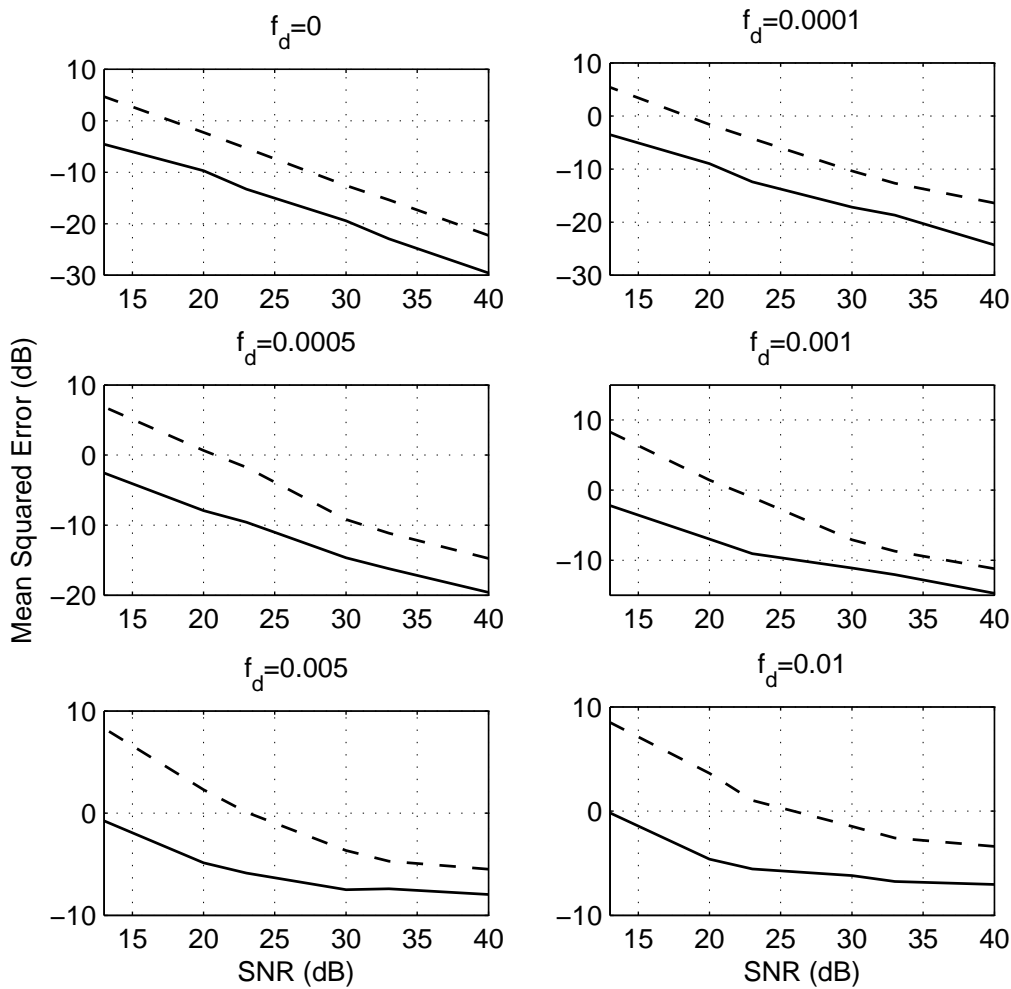}
\end{center}
\caption{MSE of RLS and SPARLS vs. SNR for $f_d = 0, 0.0001, 0.0005, 0.001, 0.005$ and $0.01$. The solid and dashed lines correspond to SPARLS and RLS, respectively.}
\end{figure}

\begin{figure}
\begin{center}
    \includegraphics[width=6in]{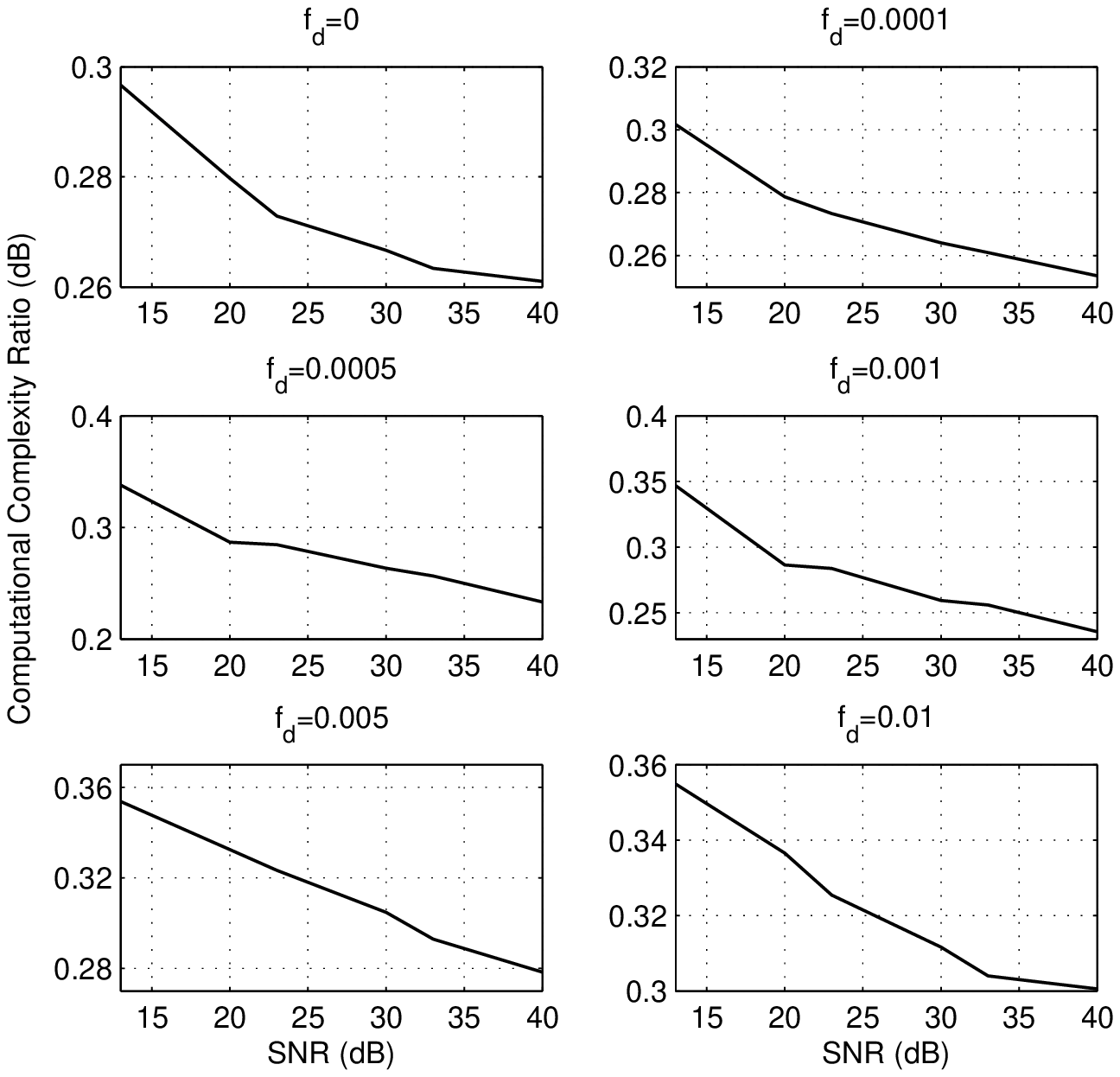}
\end{center}
\caption{CCR vs. SNR for $f_d = 0, 0.0001, 0.0005, 0.001, 0.005$ and $0.01$.}
\end{figure}



%

\section{Conclusion}\label{conclusion}

We have developed a Recursive $\mathcal{L}_1$-Regularized Least Squares (SPARLS) algorithm for the estimation of a sparse tap-weight vector in the adaptive filtering setting. The SPARLS algorithm estimates the tap-weight vector based on noisy observations of the output stream, using an Expectation-Maximization type algorithm. Simulation studies, in the context of multi-path wireless channel estimation, show that the SPARLS algorithm has significant improvement over the conventional widely-used Recursive Least Squares (RLS) algorithm, in terms of both mean squared error (MSE) and computational complexity.

\end{document}